\documentstyle[prd,aps,epsf,floats]{revtex}
\draft

\begin{document}
\twocolumn[\hsize\textwidth\columnwidth\hsize\csname
@twocolumnfalse\endcsname
\title{\vbox{\hbox to \textwidth{\normalsize\rm 
December 1995 (Final version March 1996)
\hfil Preprint MPI-PTh 95-120}
\hbox to \textwidth{\normalsize\rm To be published in Physical Review
  D \hfil E-Print ASTRO-PH/9601111}
\bigskip\bigskip
Neutrino Oscillations and the Supernova 1987A Signal}}
\author{Beat Jegerlehner, Frank Neubig, and Georg Raffelt}
\address{Max-Planck-Institut f\"ur Physik,
F\"ohringer Ring 6, 80805 M\"unchen, Germany}
\date{December 1995}
\maketitle
\begin{abstract}
We study the impact of neutrino oscillations on the interpretation
of the supernova (SN) 1987A neutrino signal by means of a
maximum-likelihood analysis. We focus on oscillations between
$\overline\nu_e$ with $\overline\nu_\mu$ or $\overline\nu_\tau$ with
those mixing parameters that would solve the solar neutrino
problem. For the small-angle MSW solution
($\Delta m^2\approx10^{-5}\,\rm eV^2$, $\sin^22\Theta_0\approx0.007$),
there are no significant oscillation effects on the Kelvin-Helmholtz
cooling signal; we confirm previous best-fit values for the
neutron-star binding energy and average spectral $\overline\nu_e$
temperature.  There is only marginal overlap between the upper end of
the 95.4\% CL inferred range of $\langle E_{\overline\nu_e}\rangle$
and the lower end of the range of theoretical predictions. Any
admixture of the stiffer $\overline\nu_\mu$ spectrum by oscillations
aggravates the conflict between experimentally inferred and
theoretically predicted spectral properties. For mixing parameters in
the neighborhood of the large-angle MSW solution
($\Delta m^2\approx10^{-5}\,\rm eV^2$, $\sin^22\Theta_0\approx0.7$)
the oscillations in the SN are adiabatic, but one needs to include the
regeneration effect in the Earth which causes the Kamiokande and
IMB detectors to observe different $\overline\nu_e$ spectra. For the
solar vacuum solution ($\Delta m^2\approx10^{-10}\,\rm eV^2$,
$\sin^22\Theta_0\approx1$) the oscillations in the SN are
nonadiabatic; vacuum oscillations take place between the SN and the
detector. If either of the large-angle solutions were borne out by the
upcoming round of solar neutrino experiments, one would have to
conclude that the SN~1987A $\overline\nu_\mu$ and/or $\overline\nu_e$
spectra had been much softer than predicted by current treatments of
neutrino transport.
\end{abstract}
\vskip2pc]


\narrowtext


\section{Introduction}

Neutrino oscillations can modify the characteristics of the neutrino
signal from a supernova (SN), in particular if matter effects are
included \cite{MS}. After the observation of the SN~1987A neutrinos by
the Kamiokande \cite{Kam} and IMB \cite{IMB} detectors many authors
\cite{Neutronization} discussed the impact of matter-induced
oscillations on the prompt $\nu_e$ burst because the first event at
Kamiokande had been observed in the forward direction, allowing for an
interpretation in terms of $\nu_e$-$e$ scattering. If this
interpretation were correct one could exclude a large area of mixing
parameters where the MSW effect in the SN envelope would have rendered
the prompt $\nu_e$ burst unobservable.

Because a single event does not carry much statistically significant
information (the first Kamiokande event may have coincidentally
pointed in the forward direction), a more interesting question for the
interpretation of the SN~1987A neutrino signal is the impact of
oscillations on the main $\overline\nu_e$ pulse which is detected by
the reaction $\overline\nu_e p\to n e^+$. The SN emits roughly equal
amounts of energy in (anti)neutrinos of all flavors, but with
different spectral characteristics.  Current treatments of neutrino
transport yield
\cite{Janka2}
\begin{equation}\label{mean}
\langle E_{\nu}\rangle=\cases{10{-}12\,{\rm MeV}&for $\nu_e$,\cr
14{-}17\,{\rm MeV}&for $\overline\nu_e$,\cr
24{-}27\,{\rm MeV}&for $\nu_{\mu,\tau}$ and
$\overline\nu_{\mu,\tau}$,} \label{E001}
\end{equation}
i.e.\ $\langle E_{\nu_e}\rangle\approx
\frac{2}{3}\langle E_{\overline\nu_e}\rangle$ and
$\langle E_{\nu}\rangle\approx\frac{5}{3}
\langle E_{\overline\nu_e}\rangle$ for the other flavors. A partial
conversion between, say, $\overline \nu_\mu$'s and $\overline\nu_e$'s
due to oscillations
would ``stiffen'' the $\overline\nu_e$ spectrum observable at Earth
\cite{Wolfenstein,SSB}. (We will always take
$\overline\nu_e$-$\overline\nu_\mu$ oscillations to represent either
$\overline\nu_e$-$\overline\nu_\mu$ or
$\overline\nu_e$-$\overline\nu_\tau$ oscillations.)
Within a plausible range of progenitor star
masses and depending on the equation of state, numerical computations
yield
\begin{equation}
E_{\rm b}= 1.5{-}4.5\times10^{53}\,{\rm erg}
\label{E001x}
\end{equation}
for the total amount of binding energy \cite{Janka1}. It is almost
entirely released in the form of neutrinos.

The expected average SN~1987A $\overline\nu_e$ energy implied by the
detected signal is about $9{-}10\,\rm MeV$, with a 95.4\% confidence
interval reaching up to $14\,\rm MeV$ in some analyses
\cite{Loredo89,Loredo95}, i.e.\ barely up to the lower end of
the theoretical predictions quoted in Eq.~(\ref{E001}). If a partial
swap $\overline\nu_e\leftrightarrow\overline\nu_\mu$ had occurred, the
expected $\overline\nu_e$ energies should have been lower, causing an
even larger strain between measured and predicted $\overline\nu_e$
energies. For an ``inverted'' mass matrix with $m_{\nu_e}>m_{\nu_\mu}$
the $\overline\nu_e$-$\overline\nu_\mu$ oscillations would have been
resonant and thus nearly complete for a large range of mixing
parameters. Therefore, such inverted-mass schemes are likely excluded
on the basis of the SN~1987A data \cite{SSB,RS}.

If the mass hierarchy is ``normal'' with $m_{\nu_e}<m_{\nu_\mu}$,
oscillations in the antineutrino sector are significant only for
large mixing angles which are often thought to be unlikely. Therefore,
in the original analyses of the SN~1987A neutrinos, little attention
has been paid to antineutrino oscillations.

Since then much progress has been made with the observation of solar
neutrinos in four experiments which all report a deficit and thus
point to oscillations.  While it remains uncertain if the solar
neutrino deficits are indeed caused by oscillations, it has become
clear that there is no simple ``astrophysical solution.'' If the
oscillation interpretation is adopted there remain three islands in
the $\sin^22\Theta_0$-$\Delta m^2$-plane (vacuum mixing angle
$\Theta_0$) where the results from all experimental measurements of
the solar neutrino flux are consistently explained, namely the
``vacuum solution'' with $\Delta m^2$ near $10^{-10}\,\rm eV^2$ and
nearly maximum mixing \cite{KP}, the ``small-angle MSW solution'' with
$\Delta m^2$ around $10^{-5}\,\rm eV^2$ and $\sin^22\Theta_0\approx
0.007$, and the ``large-angle MSW solution'' with about the same
$\Delta m^2$ and $\sin^22\Theta_0$ in the neighborhood of $0.7$
\cite{HataHaxton}. It will turn out that if one of the large-angle
solutions would be borne out by one of the forthcoming experiments
Superkamiokande, SNO, or BOREXINO, then a significant impact on the
interpretation of the SN~1987A signal could not be avoided.

In a recent study, Smirnov, Spergel, and Bahcall \cite{SSB} found that
the large-angle solutions were essentially excluded by the SN~1987A
data because of the ``stiffened'' spectra they would have caused at
the detectors. However, this conclusion relies heavily on theoretical
predictions for the spectral properties of a SN neutrino signal.
Kernan and Krauss \cite {KK}, on the other hand, arrive at the
opposite conclusion, namely that a significant oscillation effect was
actually favored by the data. Of course, they discard certain
theoretical predictions for the signal characteristics.  Smirnov,
Spergel, and Bahcall have performed a joint analysis for the
Kamiokande and IMB detectors.  However, in the neighborhood of the
large-angle MSW solution, matter-induced oscillations in the Earth are
important. They cause a different amount of ``regeneration'' of the
oscillations on the neutrino path through the Earth which was 3900 and
$8400\,\rm km$ for the Kamiokande and IMB detectors, respectively,
which thus would have observed different $\overline\nu_e$ spectra
\cite{Smirnov}.  Kernan and Krauss, on the other hand, have only
considered nonadiabatic oscillations which restrict the validity of
their analysis to $\Delta m^2\alt 10^{-10}\,\rm eV^2$, thus ignoring
the important case of the large-angle MSW solution.

Therefore, we presently reexamine the impact of large-angle neutrino
oscillations on the SN~1987A signal interpretation. If neutrino
oscillations between $\overline\nu_e$ and another flavor occur at all
with a large mixing angle, the mixing parameters probably correspond
to those solving the solar neutrino problem. Therefore, we focus on
mixing parameters in the neighborhood of the large-angle MSW solution
and of the vacuum solution of the solar neutrino problem. We will
assume thermal neutrino spectra with different temperatures for the
$\overline\nu_\mu$'s and $\overline\nu_e$'s. We will then perform a
maximum-likelihood analysis for the neutrino temperature and total
emitted energy.

In Sect.~II we discuss the assumed primary neutrino spectra and their
modification by oscillations. Sect.~III is devoted to our statistical
methodology and Sect.~IV to detailed numerical results. In Sect.~V we
summarize our findings.


\section{Neutrino Spectra}

\subsection{Primary Spectra}

The most detailed statistical analysis of the SN~1987A neutrino signal
has been performed in the papers by Loredo and Lamb
\cite{Loredo89,Loredo95} where one of the main goals was to estimate
the Kelvin-Helmholtz cooling time scale of the newly formed neutron
star, and to derive limits on the $\overline\nu_e$ mass from the
absence of pulse dispersion effects. Therefore, the time structure of
the neutrino signal was crucial; it had to be parametrized in terms of
a variety of cooling models. In our study, on the other hand, we will
focus on the spectral characteristics of the neutrino fluence
(time-integrated flux) and their modification by oscillations.
Because we will need to vary neutrino mass differences and mixing
angles, the overall number of parameters would get out of hand if we
were to analyse the time structure of the burst together with neutrino
oscillation effects.

Numerical simulations \cite{numericalmodels} and an analytic argument
\cite{Janka95} indicate an approximate equi\-partition of the energy
emitted in different (anti)neutrino species with different
time-averaged energies as quoted in Eq.~(\ref{E001}). The detailed
spectral shape, however, is not well known. Monte-Carlo studies of
neutrino transport \cite{Janka89} indicate that the instantaneous
neutrino spectra are ``pinched,'' meaning that their low- and
high-energy parts are suppressed relative to a Maxwell-Boltzmann
spectrum of the same average energy. Usually the instantaneous spectra
are expressed in the form \cite{Janka89}
\begin{equation}
f(E,t)\propto \frac{E^2}{e^{E/T-\eta}+1}\,,
\label{E002}
\end{equation}
where $\eta$ is an effective degeneracy parameter. Both $T$ and $\eta$
are functions of time. It must be stressed that the ``pseudo
degeneracy parameter'' 
$\eta$ for $\nu_\mu$
and $\nu_\tau$ is the same as that for $\overline\nu_\mu$ and 
$\overline\nu_\tau$, in contrast with the degeneracy parameter of a
real Fermi-Dirac distribution which has the opposite sign for
antineutrinos relative to neutrinos. Therefore, Eq.~(\ref{E002})
is a somewhat arbitrary two-parameter representation of the neutrino
spectra which allows one to fit two of their moments, for example
$\langle E\rangle$ and $\langle E^2\rangle$. Janka and Hillebrandt
\cite{Janka89} found that throughout the emission process $\eta$
decreases from about 5 to 3 for $\nu_e$, from about 2.5 to 2 for
$\overline\nu_e$, and from about 2 to 0 for $\nu_{\mu,\tau}$ and
$\overline\nu_{\mu,\tau}$.

The time-integrated spectrum, however, need not be pinched. We
characterize it by the moments $\langle E\rangle$ and
$\langle E^2\rangle$, and call it ``pinched'' if the ratio
$\langle E^2\rangle/\langle E\rangle^2$ is smaller than for the
Maxwell-Boltzmann case, ``anti\-pinched'' otherwise. As a simple
example we consider a cooling model where neutrinos are emitted from a
neutrino sphere with a fixed radius and an exponentially decreasing
effective temperature.  If the instantaneous spectra are of the form
Eq.~(\ref{E002}) with a fixed $\eta$, then the time-integrated
spectrum is pinched for $\eta\agt1.7$ and antipinched for
$\eta\alt1.7$.  For $\eta\approx1.7$ it is approximately of the
Maxwell-Boltzmann form.

An exponential cooling model is, of course, very simplistic. In a real
SN the $\overline\nu_e$ temperature will initially rise, and may stay
approximately constant for some time, while the effectively radiating
surface shrinks quickly within the first second.  Still, the
exponential cooling example illustrates that a thermal
Maxwell-Boltzmann spectrum may be a relatively good approximation for
the time-integrated spectrum because of the compensating effects
between instantaneous pinching and the superposition of different
spectra in the course of the protoneutron star's cooling history.
Certainly, there is no reason to expect the time-integrated spectrum
to be of the form Eq.~(\ref{E002}). This parametrization does not
allow one to describe anti\-pinched spectra, only pinched ones.

For the rest of this study we will make the simplifying assumption
that the time-integrated spectra are described by the
Maxwell-Boltzmann form
\begin{equation}
F(E) = \int_{0}^{\infty}dt\, f(E,t) \propto E^2\,e^{-E/T}
\label{E003}
\end{equation}
with a different effective temperature for $\overline\nu_e$ and
$\overline\nu_\mu$.  These ``temperatures'' are parameters
which characterize the time-integrated spectra by virtue of
$T\equiv\frac{1}{3}\langle E\rangle$ and thus do not exactly
correspond to a physical temperature at the neutron star.


\subsection{Modification by Oscillations}

In the Kamiokande and IMB detectors, SN neutrinos are almost
exclusively detected by the reaction $\overline\nu_e p\to n e^+$ where
the final-state positron is measured by its Cherenkov emission of
photons. If neutrinos do not mix, their fluence
$F_{\overline\nu_e}(E)$ relevant for the detection process is
identical with the pimary $\overline\nu_e$ spectrum
$F_{\overline\nu_e}^0(E)$ emitted from the SN. In the presence of
$\overline\nu_e\leftrightarrow\overline\nu_\mu$ oscillations, on the
other hand, each primary $\overline\nu_\mu$ arrives with a probability
$p$ in the $\overline\nu_e$ flavor state at the detector, while each
primary $\overline\nu_e$ arrives as $\overline\nu_e$ with the
``survival probability'' $1-p$ so that
\begin{equation}\label{pf}
F_{\overline\nu_e}=(1-p)\,F_{\overline\nu_e}^0
+p\,F_{\overline\nu_\mu}^0.
\end{equation}
This incoherent superposition of the individual flavor fluxes is
justified by the incoherent neutrino 
emission from different regions in the
star and by different processes~\cite{SSB}.

The ``permutation factor'' $p$ is in general a function of the
neutrino energy $E$, the mass difference $\Delta m^2$, and the vacuum
mixing angle $\Theta_0$. In addition, it is important to note that the
neutrinos are produced in a region of high matter density.  The
effective mixing angle in a medium is given by the well-known formula
\begin{equation}\label{mangle}
\tan2\Theta=\frac{\sin2\Theta_0}
{\cos2\Theta_0\mp\rho/\rho_{\rm res}}\,,
\end{equation}
where $\Theta_0$ is the vacuum mixing angle, $\rho$ the matter
density, and the upper sign refers to $\nu$, the lower to
$\overline\nu$. The ``resonance density'' is defined by
\begin{equation}\label{resonancedensity}
\rho_{\rm res} \equiv
\frac{m_{N}\Delta m^{2}}{2\sqrt{2}\,G_{\rm F}Y_{e}E},
\end{equation}
where $\Delta m^2=m_2^2-m_1^2$ with $m_2$ the dominant mass admixture
of $\nu_\mu$ and $m_1$ that of $\nu_e$. For neutrinos with a normal
mass hierarchy ($m_2>m_1$) the denominator in Eq.~(\ref{mangle})
vanishes for $\rho=\rho_{\rm res}\,\cos2\Theta_0$, causing maximum
mixing with $\Theta=\pi/4$ and thus a ``resonance.'' For
antineutrinos, and because we always assume a normal mass hierarchy,
the denominator of Eq.~(\ref{mangle}) is always larger than
$\cos2\Theta_0$ so that the medium mixing angle is always smaller than
the vacuum one.

For our purposes with neutrino energies $E\agt10\,\rm MeV$ and mass
differences $\Delta m^{2}\alt 10^{-3}\,\rm eV^2$ the resonance density
is of order $10^{3}\,\rm g\,cm^{-3}$ or less. With
$\rho\approx10^{12}\,\rm g\,cm^{-3}$ at the neutrino sphere, the
effective antineutrino mixing angle at the production site is
$\Theta\alt10^{-9}$, even if the vacuum mixing angle is maximal.
Therefore, the medium effects ``demix'' the antineutrinos, causing the
flavor eigenstates at the production site to coincide essentially with
the propagation eigenstates.

As the neutrinos leave the SN they propagate through a certain density
profile and ultimately reach the surrounding vacuum. The $\Delta m^2$
values corresponding to the large-angle solutions of the solar
neutrino problem are representative of two cases that need to be
distinguished for the further flavor evolution of the neutrino burst.

The simpler case is the vacuum solution for
$\Delta m^2\alt 10^{-10}\,\rm eV^2$. The propagation out of the SN
is not adiabatic so that the neutrinos emerge essentially as flavor
eigenstates which then oscillate on their way to Earth. Therefore, the
permutation factor has the form
\begin{equation}\label{pvac}
p={\textstyle{1\over2}}\,\sin^22\Theta_0.
\end{equation}
We note that $\Delta m^2\approx 10^{-10}\,\rm eV^2$ is at the
borderline for this statement to apply; for slightly larger mass
differences the detailed propagation through the SN envelope must be
taken into account \cite{SSB}.

For the large-angle solar MSW solution with
$\Delta m^2\approx 10^{-5}\,\rm eV^2$ we are in the adiabatic regime
where the neutrinos stay in a propagation eigenstate throughout their
journey out of the SN \cite{SSB}. What emerges is a flux of $m_1$
eigenstate neutrinos with the $\overline\nu_e$ spectrum, and one of
$m_2$ eigenstates with the $\overline\nu_\mu$ spectrum.

We stress that this statement applies even though the neutrinos
encounter a density discontinuity corresponding to the outward moving
shock wave which ultimately ejects the SN mantle and envelope.  At the
neutrino sphere, the propagation and flavor eigenstates coincide
because of the medium-induced demixing effect described above. When
the neutrinos encounter a density discontinuity in a medium so dense
that they are sufficiently demixed, then no significant flavor
transitions will occur even though this discontinuity violates the
adiabaticity condition. Within the first few seconds after collapse
the shock wave may reach a radius of at most a few $10^5\,\rm km$. In
typical progenitor star models the density varies approximately as
$r^{-3}$.  Initially, the neutrino sphere with a density of about
$10^{12}\,\rm g\,cm^{-3}$ is at a radius of about $100\,\rm km$.
Therefore, within the Kelvin-Helmholtz cooling phase the shock wave
may reach a density about 9 orders of magnitude smaller than the
neutrino sphere, i.e.\  a density as low as $10^3\,\rm g\,cm^{-3}$.
For $\Delta m^2\approx 10^{-5}\,\rm eV^2$ the resonance density
is about $10\,\rm g\,cm^{-3}$. Hence, during the entire
Kelvin-Helmholtz cooling phase the medium mixing angle is small when
the neutrinos encounter the shock wave. Therefore, the impact of level
crossing between the propagation eigenstates on the neutrino spectra
arriving at the detector can be neglected.

Because neutrinos with $\Delta m^2\approx 10^{-5}\,\rm eV^2$ emerge
from the SN as propagation eigenstates, no oscillations occur on the
way from the SN to Earth. Thus, we would have $p=\sin^2\Theta_0$ if
there were no further intervening matter.

However, in order to reach the Kamiokande and IMB detectors, the
neutrinos had to traverse $d_{\rm KAM}=3900\,\rm km$ and
$d_{\rm IMB}=8400\,\rm km$ of matter in the Earth, with an average
density of about $\rho_{\rm KAM}=3.4\,\rm g\,cm^{-3}$ and
$\rho_{\rm IMB}=4.6\,\rm g\,cm^{-3}$, respectively \cite{SSB}.
Therefore, the permutation factor relevant for each detector is
\cite{SSB}
\begin{equation}\label{perde} p=\sin^2\Theta_0-
\sin2\Theta\,\sin(2\Theta_0-2\Theta)\,
\sin^2(\pi d/\ell).
\end{equation}
The medium mixing angle relevant for each detector is given by
Eq.~(\ref{mangle}) with $\rho=\rho_{\rm KAM}$ or $\rho_{\rm IMB}$,
respectively, the distance in Earth is $d=d_{\rm KAM}$ or
$d_{\rm IMB}$, and the oscillation length is
\begin{equation}
\ell=\frac{4\pi E}{\Delta m^2}\,\frac{\sin 2\Theta}{\sin2\Theta_0}
\end{equation}
with the relevant medium mixing angle. For the solar vacuum solution
with $\Delta m^2\approx10^{-10}\rm\,eV^2$ the Earth effect is
unimportant.


\section{Statistical Methodology}

\subsection{Parameter Estimation and Confidence Regions}

The purpose of the present study is to estimate the parameters
$E_{\rm b}$ and $T_{\overline\nu_e}$ which characterize the neutrino
fluence from SN~1987A and to study the impact of neutrino mixing on
this estimate.  Because of the small number of SN~1987A events in the
Kamiokande and IMB detectors this task is rather delicate. One needs a
statistical estimator which is consistent and unbiased, and which
exploits the sparse data efficiently. The maximum-likelihood method
\cite{Eadie,Kendall} is particularly well suited for such problems,
i.e.\ problems where it is essential to extract the maximum possible
information from a small number of events. This method has been used
by several authors to analyse the SN~1987A neutrino signal, e.g.\
Refs.~\cite{Loredo89,Loredo95,KK,Janka89}.

The method consists of deriving the set of parameters, collectively
denoted by $\alpha$, for which the probability of producing the
observed data set, collectively denoted by $x$, becomes maximal.
The probability density as a function of $\alpha$ for producing the
observed data is called the likelihood function ${\cal L}(x,\alpha)$.
The maximum-likelihood estimation $\alpha_*$ for the true but unknown
parameter set $\alpha_0$ is implicitly defined by
\begin{equation}
{\cal L}(x,{\alpha_*}) =
\max_{\alpha \in D}\,{\cal L}(x,\alpha),
\end{equation}
where $D$ is the parameter domain.

An estimation $\alpha_*$ of the true parameters $\alpha_0$ is useful
only if one also determines a confidence region around $\alpha_*$
which contains the true parameters with a specified probability
$\beta$.  To construct this region assume that the true parameters
$\alpha_0$ are given. We can then determine the probability
distribution $P_{\alpha_0}(\alpha_*)$ of the likelihood estimator and
define a region $D_{\beta,\alpha_0}$ from the condition
$P_{\alpha_0}(\alpha_*)\ge\beta$ for $\alpha_*\in D_{\beta,\alpha_0}$.
To make it unique we additionally require that
$P_{\alpha_0}(\alpha_*)$ is larger for all $\alpha_*$ within
$D_{\beta,\alpha_0}$ than for those outside. Put another way, we
require $D_{\beta,\alpha_0}$ to be bounded by a contour of constant
$P_{\alpha_0}(\alpha_*)$. The confidence region $D^*_\beta$ can now be
defined as the region of parameters $\alpha$ for which
$\alpha_*\in D_{\beta,\alpha}$. Note that this set is in general not
equal to $D_{\beta,\alpha_*}$.

In practice, this region is difficult to calculate because finding
$D_{\beta,\alpha}$ alone requires integrating over the space of
possible observations, a task usually achieved by Monte-Carlo
sampling.  However, if $\cal L$ is Gaussian the confidence region is
given by the condition
\begin{equation}
\ln{\cal L}(x,\alpha_*)-\ln {\cal L}(x,\alpha)\le
{\textstyle \frac{1}{2}}\,\chi_\beta(k),
\end{equation}
again with the additional requirement that it should be bounded by a
contour of constant ${\cal L}$ in parameter space \cite{Eadie}.
Further, $k$ is the number of parameters which for our study will
usually be $k=2$. Note that $\chi_\beta(2)=2.3$, 4.61, and 6.17 for
$\beta=68.3\%$, $90\%$, and $95.4\%$, respectively. We stress that the
confidence regions thus determined are not exact, especially when they
are very distorted so that the parameters are strongly correlated.


\subsection{Likelihood Function}

It is not trivial to determine the likelihood function appropriate for
our problem. The primary observations of the water Cherenkov detectors
consist of the information when a given photomultiplier has fired.
This information can be used to reconstruct the event location in the
detector and the energy of the detected charged particle. For our
purposes it is probably sufficient to use the reported event energies
as the primary data set and assume that they are related to the true
positron energies by a Gaussian distribution.

In order to model the likelihood function we consider detection energy
bins $[E_i, E_i+\delta E]$ with $i=1,\ldots, N_{\rm bin}$. The
spectrum of detected energies is $n(E)$ so that the number of expected
counts in bin $i$ is to lowest order $n(E_i)\,\delta E$.  However, in
a real experiment one obtains an integer number $N_i$ of counts in a
given bin $i$. The probability for such an outcome is
\begin{equation}
P_i=\frac{[n(E_i)\,\delta E]^{N_i}}{N_i!}\,
e^{-n(E_i)\,\delta E},
\end{equation}
where the $N_i$ are the actual observations and thus represent the
data. The likelihood function is
\begin{equation}
{\cal L}=\prod^{N_{\rm bin}}_{i=1} P_i\,.
\end{equation}
This expression can be transformed to
\begin{equation}
{\cal L}=C\, e^{-\int_0^\infty n(E)dE}
\prod_{i=1}^{N_{\rm obs}} n(E_i),
\end{equation}
where $N_{\rm obs}$ is the total number of experimentally observed
events. The constant $C$ is irrelevant for the purpose of parameter
estimation and the determination of confidence regions.  For a joint
analysis of the Kamiokande and IMB detectors, the likelihood function
is the product of the likelihood functions for each detector.


\subsection{Expected Energy Spectrum}

In order to translate the $\overline\nu_e$ fluence
$F_{\overline\nu_e}(E)$ at Earth to an expected spectrum $n(E)$ of
counts we must first determine the energy spectrum of secondary
positrons in the $\overline\nu_e p\to n e^+$ reaction. Its cross
section $\sigma_{\overline\nu_e p}$ as a function of neutrino energy
$E$ is
\begin{equation}
\sigma_0\,\left(\frac{E}{m_e}\right)^2\,
\left(1-\frac{Q}{E}\right)\,\left[1-\frac{2Q}{E}
+\frac{Q^2-m_e^2}{E^2}\right]^{1/2}\!\!,
\end{equation}
where $Q=1.29\,\rm MeV$ is the neutron-proton mass difference, $m_e$
the electron mass, and $\sigma_0=2.295\times10^{-44}\,\rm cm^2$. We
ignore Coulomb and radiative corrections as well as neutron recoils.
Therefore, the positron spectrum in the detector is
\begin{equation}
n_+(E)=\frac{N_p}{4\pi D^2}\,\sigma_{\overline\nu_e p}(E+Q)
\,F_{\overline\nu_e}(E+Q),
\end{equation}
where $D=50\,\rm kpc$ is the distance to the SN and $N_p$ the number
of target protons in a given detector, namely $1.43 \times 10^{32}$
for Kamiokande and $4.55 \times 10^{32}$ for IMB.

The positron spectrum $n_+(E)$ produced in the detector is not
identical with the spectrum $n(E)$ of events that one expects to
detect. The reported energy $E_{\rm det}$ for an event is
reconstructed from the number of photomultipliers that have
been triggered by the Cherenkov light of the positrons produced
in the detector. Because this involves a Poissonian process, a certain
number of active photomultipliers corresponds to a range of possible
positron energies $E_+$ that may have caused this event. Moreover,
there is an $E_+$ dependent efficiency curve $\eta_0(E_+)$ that a
given positron will trigger the detector at all.  While this function
is essentially a step function for the Kamiokande detector, it is
fairly nontrivial for IMB where about a quarter of the
photomultipliers were not operational at the time of SN~1987A due to
a failed power supply.

The spectrum of possible reconstructed event energies $E_{\rm det}$
that may be attributed to a true positron energy $E_+$ is not
universal throughout the detector; there are nontrivial geometry
effects. Still, we use a universal distribution for the probability of
finding $E_{\rm det}$ if the true energy was $E_+$,
\begin{equation}\label{prob}
P(E_{\rm det},E_+)=\frac{1}{\sqrt{2\pi}\,\sigma(E_+)}\,
\exp\left(-\frac{(E_+-E_{\rm det})^2}{2\sigma^2(E_+)}\right).
\end{equation}
Motivated by the Poissonian nature of the detection process we
approximate the energy-dependent dispersion by
\begin{equation}
\sigma(E_+)=\sqrt{E_\sigma E_+}.
\end{equation}
For each detector we fit $E_\sigma$ from the uncertainties of the
reported experimental event energies \cite{Kam,IMB}. We find that a
good approximation is $E_\sigma=0.75\,\rm MeV$ for Kamiokande and
$1.35\,\rm MeV$ for IMB.

Instead of using a universal function for $P(E_{\rm det},E_+)$ we
could have used the reported experimental errors $\sigma_i$ for each
event. This procedure would leave our results almost unchanged while
causing complications for the definition of an overall detector
efficiency curve below.

In both detectors a trigger threshold for the minimum number of
photomultipliers was used in order to attribute a given event to an
external signal rather than to background. This corresponds to a lower
$E_{\rm det}$ threshold of $E_{\rm cut}=7.5\,\rm MeV$ for Kamiokande
and $19\,\rm MeV$ for IMB. The published trigger efficiency curves
$\eta(E_+)$ are thus to be interpreted as
\begin{equation}
\eta(E_+)=\eta_0(E_+)\int_{E_{\rm cut}}^\infty dE_{\rm det}\,
P(E_{\rm det},E_+),
\label{E010}
\end{equation}
where $\eta_0(E_+)$ represents efficiency reductions from other causes
such as geometry and dead-time effects. 

\begin{figure}
\centering\leavevmode
\epsfxsize=2.8in
\epsfbox{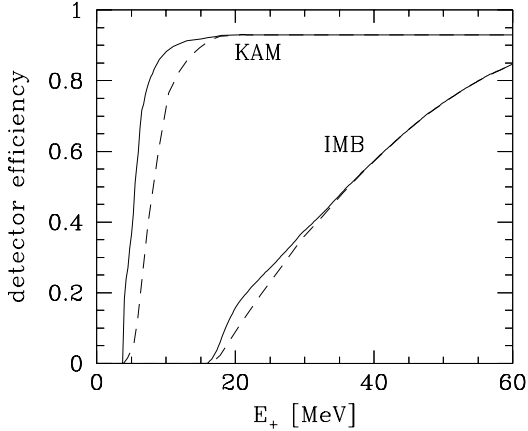}
\smallskip
\caption[...]{Efficiency curves for Kamiokande and IMB. A 13\%
dead-time effect for IMB is not included.
The $\eta$ curves (dashed) represent the overall efficiencies
published in Refs.~\cite{Kam,IMB} while the $\eta_0$ curves (solid)
are corrected according to Eq.~(\ref{E010}) for the ``smearing-out'' 
of $E_{\rm det}$ relative to the positron energy~$E_+$.
\label{Fig1}}
\end{figure}

In Fig.~\ref{Fig1} we show
$\eta(E_+)$ and $\eta_0(E_+)$ for both Kamiokande and IMB where for
the latter detector a 13\% dead-time effect is not taken into account
in the efficiency curve. For Kamiokande, $\eta_0(E_+)$ is essentially
constant down to the threshold, revealing that the efficiency curve
$\eta(E_+)$ is dominated by the trigger threshold and by the
Poissonian nature of the detection process. For IMB, on the other
hand, there is a significant geometrical efficiency modification.

The expected spectrum of detected energies is thus related to the
actual positron spectrum by
\begin{equation}
n(E_{\rm det})=\int_0^\infty dE_+\,P(E_{\rm det},E_+)\,
\eta_0(E_+)\,n_+(E_+)
\end{equation}
for $E_{\rm det}\ge E_{\rm cut}$, and $n(E_{\rm det})=0$ otherwise.
With this result we are armed to perform the maximum likelihood
analysis.


\subsection{Detector Background}

The statistical analysis described above ignores the detector
background, i.e.\ the fact that any event ascribed to the SN burst can
also be due to background, and conversely, any event attributed to
background can have been caused by the SN burst. In Loredo and Lamb's
analyses \cite{Loredo89,Loredo95} the background spectrum was included
in the expected event rate. Events much earlier or much later than the
main burst are automatically discriminated against and thus do not
overdominate the low-energy part of the expected event distribution.
Without the possibility to discriminate against background events by
the temporal relationship to the main burst we must use the cut
represented by the energy threshold $E_{\rm cut}$. We stress that
including the background as in Loredo and Lamb's analyses does not
cause a large modification of the implied SN binding energy and
neutrino temperature.


\section{Numerical Results}

\subsection{No Mixing}

For comparison with previous work we begin our max\-imum-likelihood
analysis with the case of no neutrino mixing. We search for the
best-fit SN binding energy $E_{\rm b}$ and the effective
$\overline\nu_e$ temperature $T_{\overline{\nu}_e}$ which
characterizes the assumed Maxwell-Boltzmann $\overline\nu_e$ spectrum
of the time-integrated flux by virtue of
$\langle E_{\overline\nu_e}\rangle=3T_{\overline{\nu}_e}$. We
assume equipartition of the released SN energy between all
(anti)neutrino species so that $E_{\rm b}$ is given by six times the
inferred total energy emitted in $\overline\nu_e$'s.

\begin{figure}[b]
\centering\leavevmode
\epsfxsize=2.8in
\epsfbox{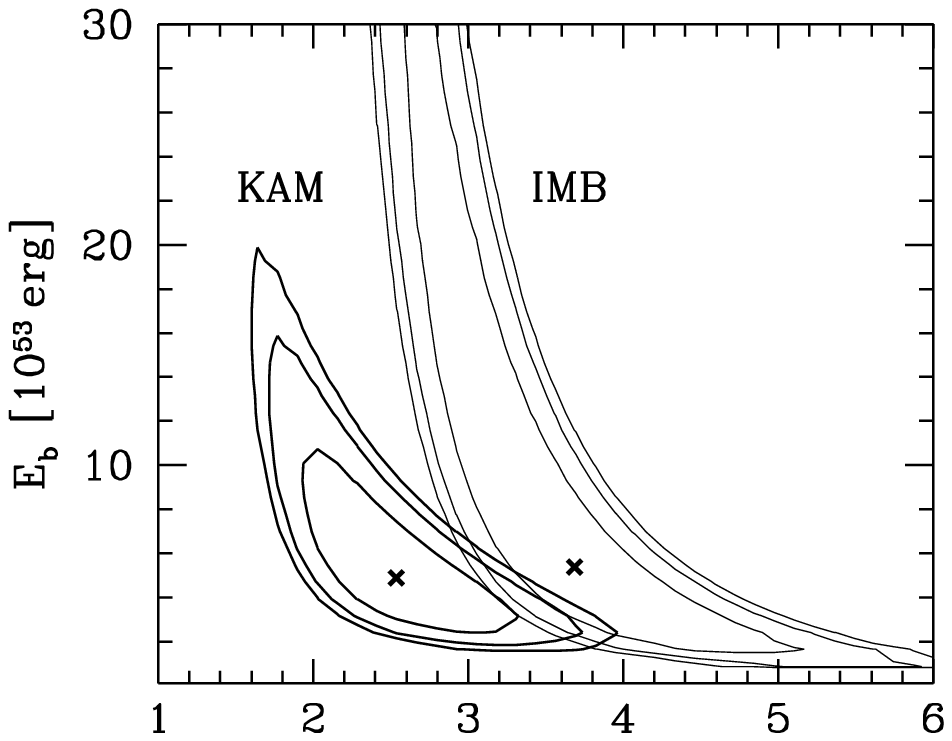}

\centering\leavevmode
\epsfxsize=2.8in
\epsfbox{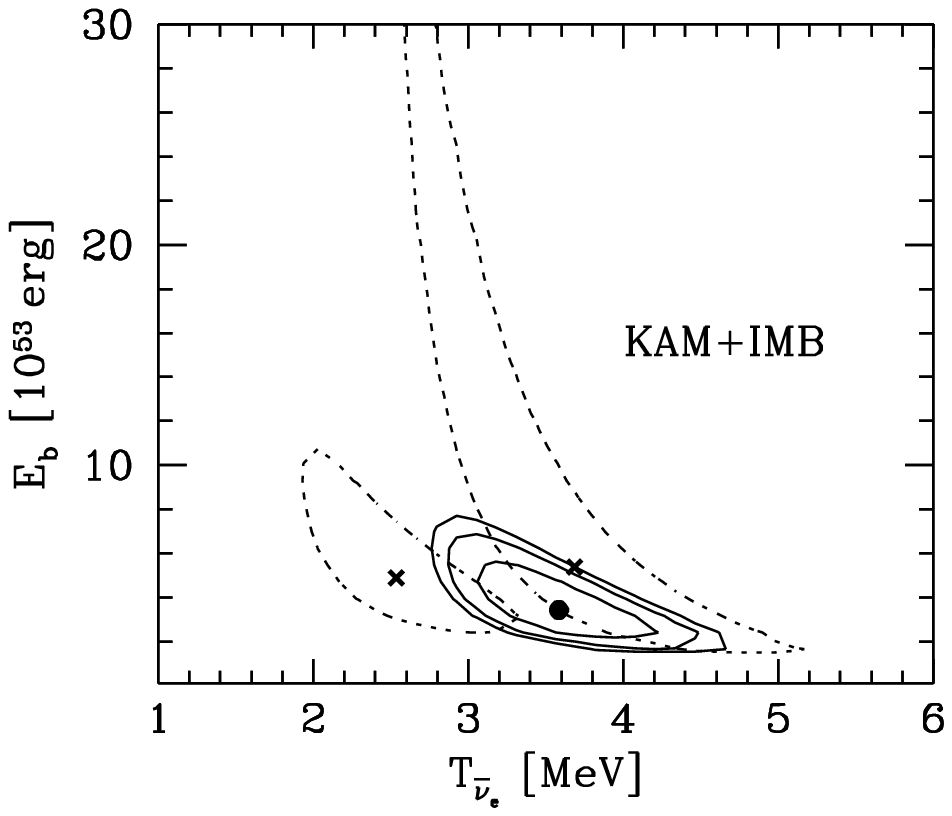}
\smallskip
\caption[...]{Contours of constant likelihood which correspond to
68.3\%, 90\%, and 95.4\% confidence regions, and best-fit values for
$T_{\overline{\nu}_e}$ and $E_{\rm b}$. {\it Upper panel:} Kamiokande
and IMB separately. {\it Lower panel:} Joint analysis. Dashed lines
mark the 68.3\% confidence regions of the separate fit.
\label{Fig2}}
\end{figure}

In Fig.~\ref{Fig2} we show the contours of constant likelihood in the
$T_{\overline{\nu}_e}$-$E_{\rm b}$-plane which correspond to 68.3\%,
90\%, and 95.4\% confidence regions, respectively, and the best-fit
values for $T_{\overline{\nu}_e}$ and $E_{\rm b}$. In the upper panel
we show the results from separate analyses for the Kamiokande and IMB
detectors, in the lower panel from a joint analysis. Our best-fit
values for the Kamiokande detector are
$T_{\overline\nu_e}=2.5\,{\rm MeV}$ and
$E_{\rm b}=4.9 \times 10^{53}\,{\rm erg}$ while for IMB
they are $3.7\,{\rm MeV}$ and $5.4\times10^{53}\,{\rm erg}$,
respectively. With the Kamiokande best-fit spectrum we find 11
neutrino events for Kamiokande and about 1 for IMB. Conversely, the
IMB best-fit spectrum yields about 24 Kamiokande and 8 IMB events.

While the overlap between the separate confidence contours is somewhat
marginal, it is sufficient to allow for a joint analysis.  The joint
best-fit values are $T_{\overline\nu_e}=3.6\,{\rm MeV}$ and
$E_{\rm b}=3.4 \times 10^{53}\,{\rm erg}$. These best-fit parameters
as well as the event numbers and average event energies corresponding
to them are summarized in Tab.~\ref{tab1}.

Our results differ somewhat from those of Janka and Hillebrandt
\cite{Janka89} in that these authors find more restrictive confidence
contours. We believe that the difference is caused by their use of a
simplified likelihood function where $E_{\rm det}$ is identified with
$E_+$ without allowing for a smearing-out effect, and by their use of
a Gaussian rather than a Poissonian modulation of the detection
process.

The inferred neutron-star binding energy agrees well with theoretical
expectations of $E_{\rm b}=1.5{-}4.5\times 10^{53}\,{\rm erg}$. The
best-fit $\langle E_{\overline\nu_e}\rangle$, however, is rather low
compared with the range of theoretical predictions quoted in
Eq.~(\ref{E001}); only the 95.4\% confidence region slightly
touches the predicted range.

\begin{table}
\centering\leavevmode
\begin{tabular}[t]{llrrrr}
\noalign{\vskip2pt}
&&Data&\multicolumn{3}{c}{Best-Fit for Mixing}\\
\noalign{\vskip 2pt}
&&&None&Vacuum&Adiabatic\\
\noalign{\vskip 4pt}
\hline
\noalign{\vskip 4pt}
$\sin^22\Theta_0$        &&---&---&0.58    &1\\
$\Delta m^2~[{\rm eV}^2]$&&---&---&---&$3.2\times10^{-6}$\\
$E_{\rm b}$ [$10^{53}\,{\rm erg}$]&&---&  3.4  &  5.6 &   9.6  \\
$T_{\overline\nu_e}$  [MeV]       &&---&  3.6  &  2.1 &   1.9  \\
$\Delta\ln({\cal L}_{\rm max})$   &&---&  0.0  &  1.3 &   3.7  \\
$N_{\rm events}$&KAM               & 11& 14.5  & 14.6 &  13.1  \\
   &IMB                            &  8&  4.5  &  4.4 &   5.8  \\
$\langle E_{\overline\nu_e}\rangle$ [MeV]
   &KAM                            &15.4& 19.9  & 19.3 &  17.1  \\
   &IMB                            &32.0& 32.6  & 34.5 &  33.7  \\
\noalign{\vskip2pt}
\end{tabular}
\medskip
\caption[...]{Best-fit values for the SN~1987A parameters for three
neutrino mixing scenarios with a relative $\overline\nu_\mu$
temperature $\tau=T_{\overline\nu_\mu}/T_{\overline\nu_e}= 2.0$
each. The expected event numbers and energies result from the joint
analysis for the Kamiokande and IMB detector. The maximum likelihood
$\Delta\ln({\cal L}_{\rm max})$ is relative to the no-mixing case.
The case of vacuum oscillations corresponds to
$\Delta m^2\alt 10^{-10}\,{\rm eV}^2$ but is otherwise independent of
the mass difference.
\label{tab1}}
\end{table}


\begin{figure}[b]
\centering\leavevmode
\epsfxsize=2.8in
\epsfbox{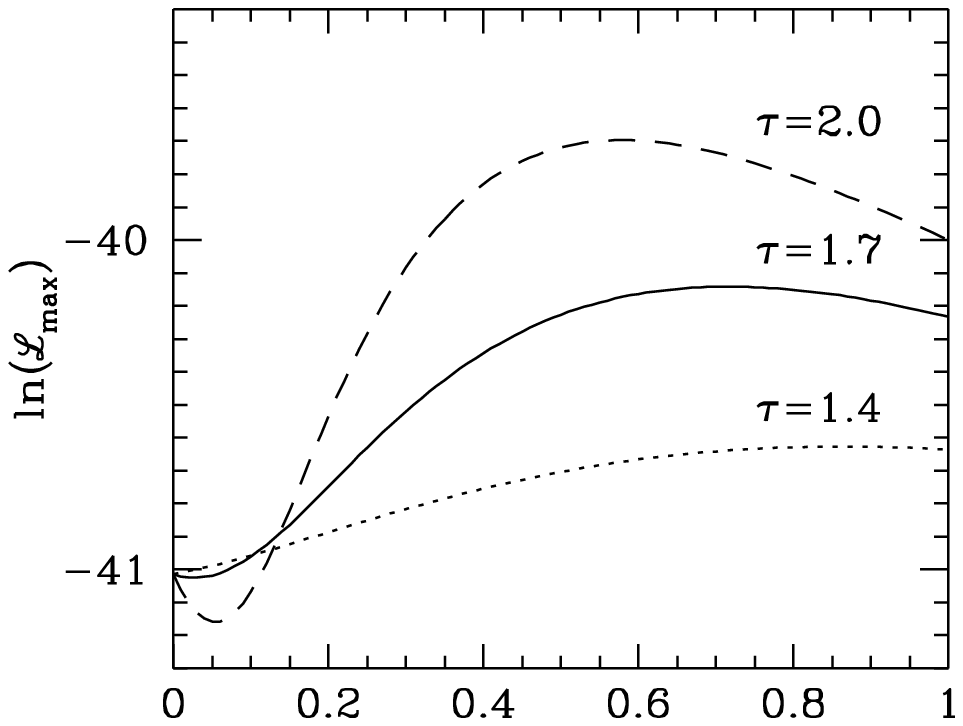}

\centering\leavevmode
\epsfxsize=2.8in
\epsfbox{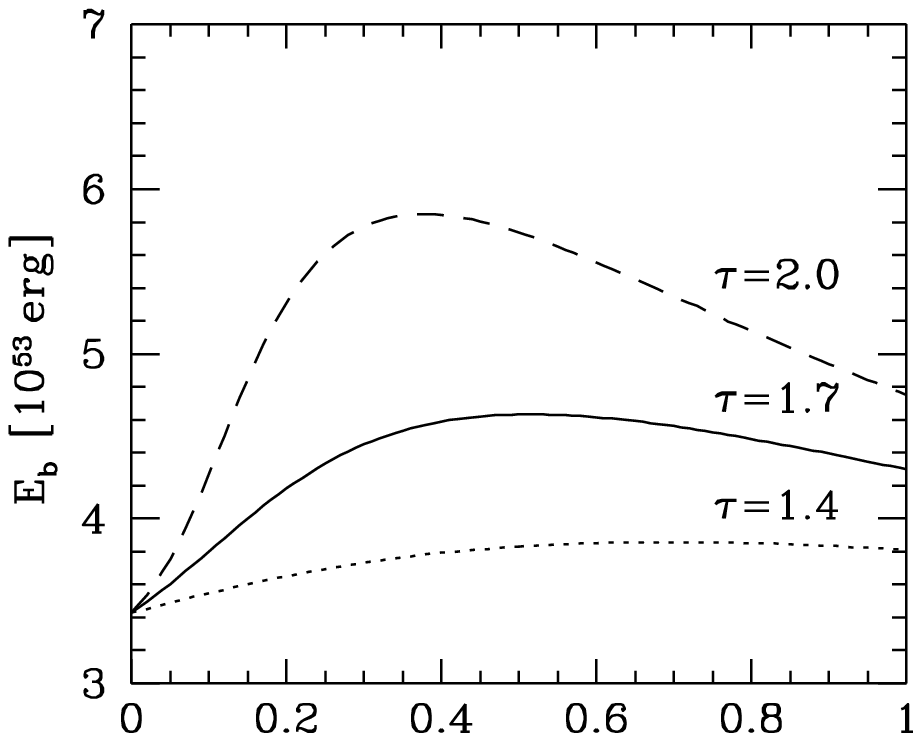}

\centering\leavevmode
\epsfxsize=2.8in
\epsfbox{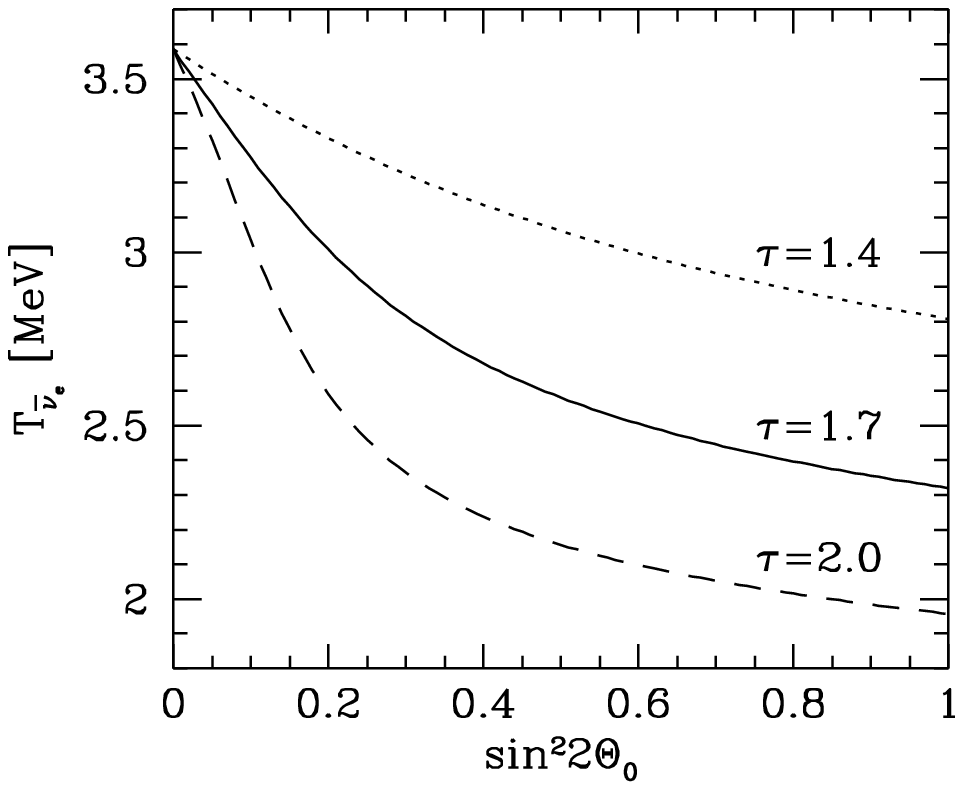}
\smallskip
\caption[...]{Maximum likelihood, binding energy, and $\overline\nu_e$
temperature as functions of the vacuum mixing angle. The
$\overline\nu_\mu$ temperature is given by
$T_{\overline\nu_{\mu}}=\tau\,T_{\overline\nu_e}$ with the indicated
$\tau$ values.
\label{Fig3}}
\end{figure}

\subsection{Vacuum Oscillations}

Next, we study the case of vacuum oscillations which is relevant for
small neutrino mass differences ($\Delta m^2\alt
10^{-10}\,\rm eV^2$). The swap probability $p$ is given by the simple
formula Eq.~(\ref{pvac}) which depends only on the vacuum mixing
angle so that no explicit dependence on $\Delta m^2$ obtains. In
analogy to the $\overline\nu_e$'s we describe the time-integrated
$\overline\nu_{\mu}$ flux by a Maxwell-Boltzmann spectrum with the
same total energy, but with a higher effective temperature
$T_{\overline\nu_{\mu}} = \tau\,T_{\overline\nu_e}$, where the factor
$\tau$ is predicted to lie in the range 1.4--2.0.

\eject

We begin by performing the maximum-likelihood analysis for a fixed
vacuum mixing angle and a fixed $\tau$-factor while allowing
$E_{\rm b}$ and $T_{\rm \overline\nu_e}$ to float.  In Fig.~\ref{Fig3}
we show as a function of $\sin^22\Theta_0$ the maximum likelihood and
the best-fit $E_{\rm b}$ and $T_{\rm \overline\nu_e}$. We show these
curves for $\tau=1.4$, 1.7, and~2.0.

For $\tau=2.0$ our results agree well with those of Kernan and Krauss
\cite{KK}. The maximum-likelihood curve has a maximum for
$\sin^22\Theta_0\approx0.5$ so that a relatively large mixing angle
appears to be favored by the data. The inferred SN parameters and
expected detector signals for this case are summarized in
Tab.~\ref{tab1}. In Fig.~\ref{Fig3} the inferred best-fit binding
energy is greater for large mixing angles compared to the no-mixing
case, while the best-fit spectral temperature is a monotonically
decreasing function of $\sin^22\Theta_0$. For $\sin^22\Theta_0\agt0.5$
and $\tau = 2.0$ the best-fit $\langle E_{\overline\nu_e}\rangle$ is
below $6\,{\rm MeV}$. Such a value is far below what is predicted
theoretically so that it looks like large mixing angles are difficult
to reconcile with the SN~1987A data.

\begin{figure}[hb]
\centering\leavevmode
\epsfxsize=2.8in
\epsfbox{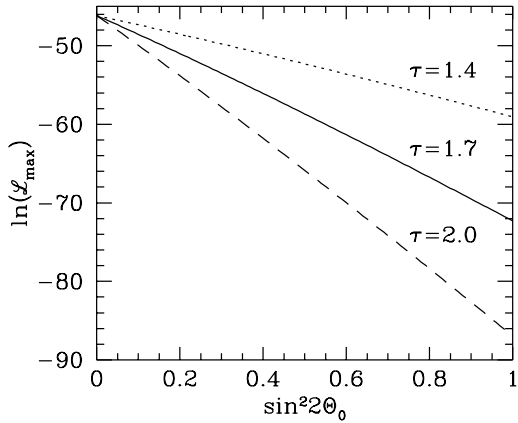}
\smallskip
\caption[...]{Likelihood for a fixed $E_{\rm b}=3\times 10^{53}\,
{\rm erg}$ and $\langle E_{\overline{\nu}_e}\rangle=14\,{\rm MeV}$ as
a function of the vacuum mixing angle. The $\overline\nu_\mu$
temperature is given by
$T_{\overline\nu_{\mu}}=\tau\,T_{\overline\nu_e}$ with the indicated
$\tau$ values.
\label{Fig4}}
\end{figure}

We can also fix the binding energy and neutrino temperature according
to theoretical predictions. Figure~\ref{Fig4} shows $\ln({\cal L})$ 
for $E_{\rm b}=3\times 10^{53}\,{\rm erg}$ and 
$\langle E_{\overline{\nu}_e}\rangle=14\,{\rm MeV}$ as a function of 
the mixing angle for several values of the relative $\overline\nu_\mu$
temperature.  The likelihood is a monotonically decreasing function of
$\sin^22\Theta_0$ so that, taking the predicted SN parameters
seriously, the best-fit mixing angle is zero, and large mixing angles
are disfavored. For $\tau=1.4$ the 95.4\% confidence interval is
$0\leq\sin^22\Theta_0\leq 0.17$.

\begin{figure}[ht]
\centering\leavevmode
\epsfxsize=2.8in
\epsfbox{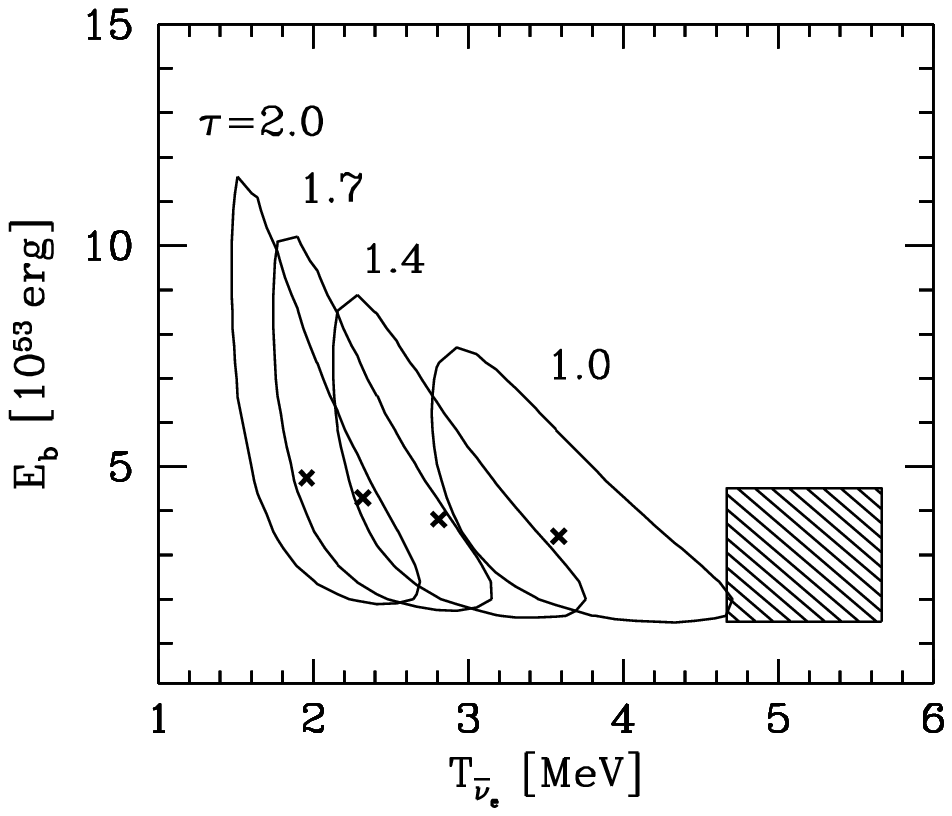}
\smallskip
\caption[...]{Best-fit values for $T_{\overline{\nu}_e}$ and
$E_{\rm b}$, and contours of constant likelihood which correspond to
95.4\% confidence regions. In each case a joint analysis between both
detectors was performed with $\sin^22\Theta_0=1$ and the indicated
relative $\overline\nu_\mu$ temperature $\tau$. The hatched region
corresponds to the theoretical predictions of Eqs.~(\ref{E001}) and
(\ref{E001x}).
\label{Fig5}}
\end{figure}

Suppose that future experiments will establish vacuum oscillations as
a solution of the solar neutrino problem. What would this imply for
the SN~1987A parameters?  To study this question we show in
Fig.~\ref{Fig5} the 95.4\% confidence contours in the
$T_{\rm \overline\nu_e}$-$E_{\rm b}$-plane for a joint analysis
between the detectors with $\sin^22\Theta_0=1$ and with $\tau=1.0$,
$1.4$, $1.7$, and $2.0$.

\eject

The 1.0 case corresponds to no mixing; the contour is identical with
that of the lower panel of Fig.~\ref{Fig2}.  The maximum
$\overline{\nu}_e$ temperature within the 95.4\% confidence region is
about $4.6\,{\rm MeV}$, corresponding roughly to the lower limit for
the range of predicted $\langle E_{\overline{\nu}_e}\rangle$ values as
given in Eq.~(\ref{mean}).

For $\tau=1.4$ the 95.4\% CL region for the $\overline\nu_e$ energies
does not overlap with theoretical predictions. Therefore, if the
vacuum solution would be borne out by future solar neutrino
experiments, one would be forced to conclude that there is a
significant problem with the predicted SN neutrino spectra and
energies.


\subsection{Adiabatic Oscillations and Earth Effect}

The most complicated case obtains if the solar neutrino problem is
solved by large-angle MSW oscillations where $\Delta m^2\approx
10^{-5}\,\rm eV^2$. The propagation out of the SN is adiabatic so that
no oscillations occur between there and the Earth, but we need to
include regeneration effects caused by the matter-induced oscillations
in the Earth. The permutation factor Eq.~(\ref{perde}) is different
for the two detectors; it is a function of the mass difference, the
vacuum mixing angle and the neutrino energy.

\begin{figure}[t]
\centering\leavevmode
\epsfxsize=2.8in
\epsfbox{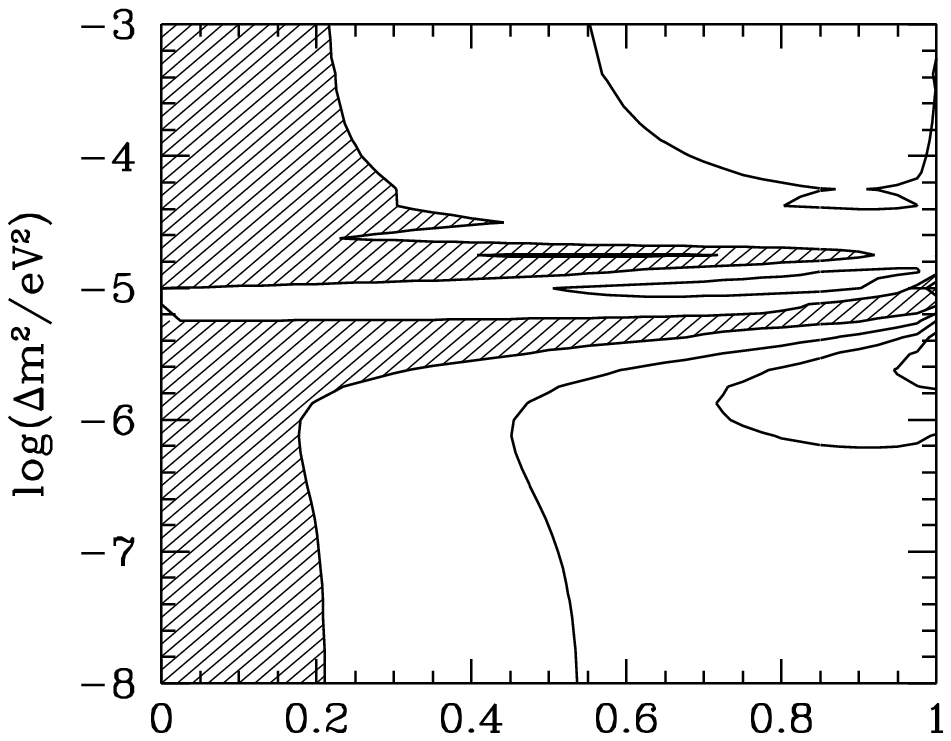}

\centering\leavevmode
\epsfxsize=2.8in
\epsfbox{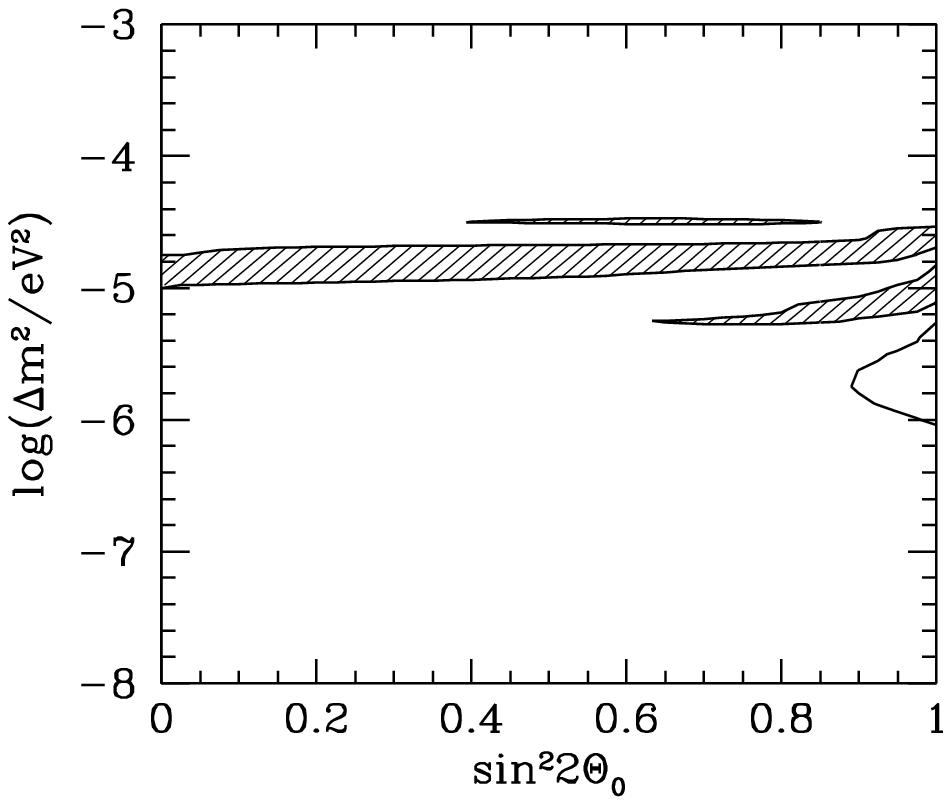}
\smallskip
\caption[...]{Contours of $\Delta\ln({\cal L}_{\rm max})$, which is
the maximum likelihood relative to the no-mixing value
$\ln({\cal L}_{\rm max})=-41.0$. The contour lines are in steps of 1.
Shaded areas correspond to $\Delta\ln({\cal L}_{\rm max})<0$, i.e.\
regions which are disfavored relative to the no-mixing case. The
relative $\overline\nu_\mu$ temperature $\tau$ was 2.0 (upper panel)
and 1.4 (lower panel).
\label{Fig6}}
\end{figure}

As in Sect.~IV.B we begin by performing the maxi\-mum-likelihood
analysis for a fixed $\Delta m^2$ and $\sin^22\Theta_0$ while allowing
$E_{\rm b}$ and $T_{\overline\nu_e}$ to float.  In Fig.~\ref{Fig6} we
show contours of $\ln({\cal L}_{\rm max})$ relative to the no-mixing
value $\ln({\cal L}_{\rm max})=-41.0$ in steps of 1.  We have used
$\overline\nu_\mu$ fluences with the same total energy as for
$\overline\nu_e$ and a relative temperature $\tau=2.0$ (upper panel)
and $\tau=1.4$ (lower panel).  The shaded areas correspond to a
negative $\Delta\ln({\cal L}_{\rm max})$ and thus to a reduced
likelihood relative to the no-mixing case. We emphasize that these
areas cannot be interpreted as being excluded even though they are
disfavored.

\begin{figure}[t]
\centering\leavevmode
\epsfxsize=2.8in
\epsfbox{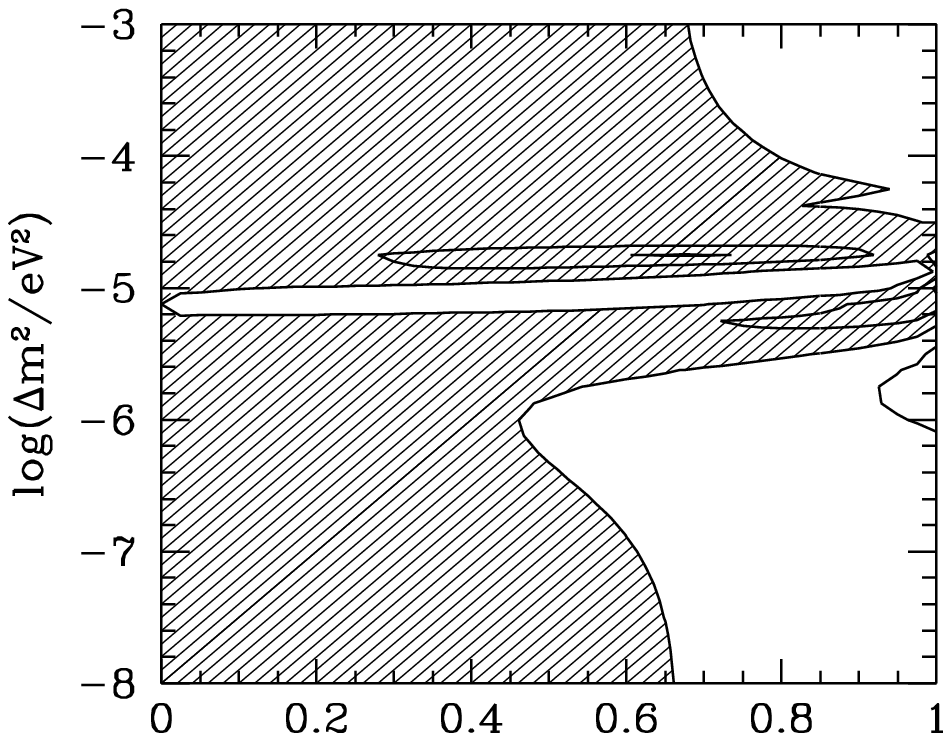}

\centering\leavevmode
\epsfxsize=2.8in
\epsfbox{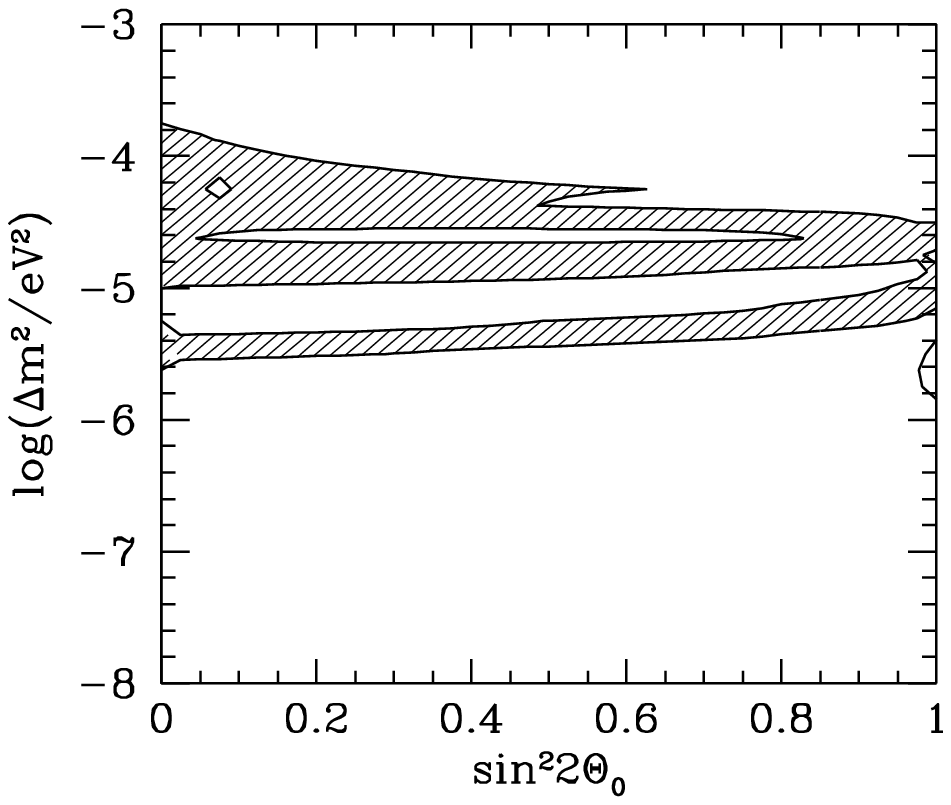}
\smallskip
\caption[...]{Contours of $\Delta\ln({\cal L}_{\rm max})$ relative to
the no-mixing case for a fixed SN binding energy
$E_{\rm b}=3\times10^{53}\,\rm erg$. The contour lines are in steps of
1. Shaded areas correspond to $\Delta\ln({\cal L}_{\rm max})<0$, i.e.\
regions which are disfavored relative to the no-mixing case. The
relative $\overline\nu_\mu$ temperature $\tau$ was 2.0 (upper panel)
and 1.4 (lower panel).
\label{Fig7}}
\end{figure}

For both $\tau=2.0$ and $1.4$ we find best-fit mixing parameters
$\sin^22\Theta_0=1$ and $\log(\Delta m^2/{\rm eV^2})\approx -5.5$. The
absolute maximum of the likelihood is
$\Delta\ln({\cal L}_{\rm max})\approx 3.7$ and $1.6$, respectively,
relative to the no-mixing case. A local maximum with
$\Delta\ln({\cal L}_{\rm max})\approx 1.4$ (0.4) is found for
$\sin^22\Theta_0\approx 0.8$ and
$\log(\Delta m^2/{\rm eV^2})\approx-5$. The largest increase of the
maximum likelihood occurs for the largest relative $\overline\nu_\mu$
temperature $\tau=2.0$. The corresponding best-fit SN parameters and
expected signal characteristics are listed in Tab.~\ref{tab1}. They
are far away from theoretical predictions so that the apparent
improvement of the likelihood is obtained at the price of a conflict
with SN theory.

\begin{figure}
\centering\leavevmode
\epsfxsize=2.8in
\epsfbox{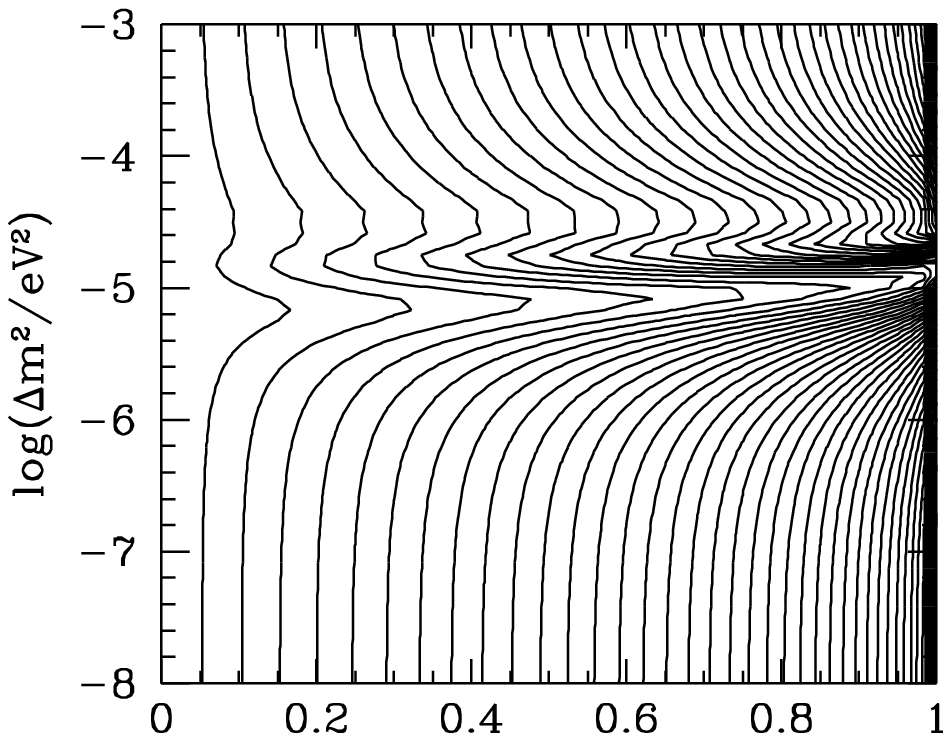}

\centering\leavevmode
\epsfxsize=2.8in
\epsfbox{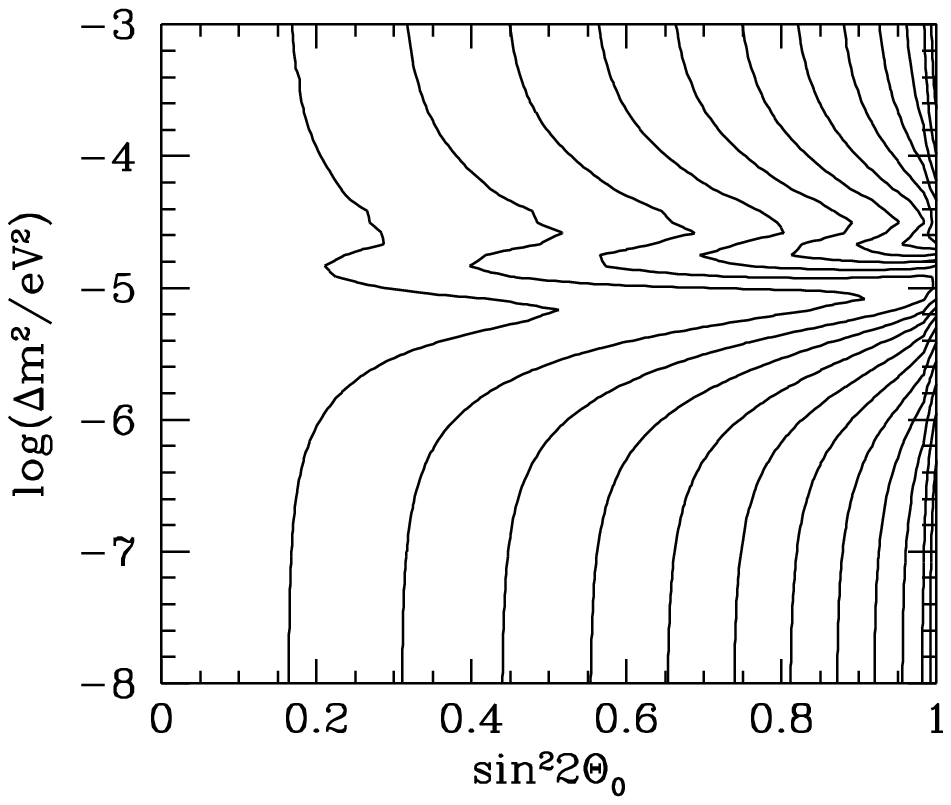}
\smallskip
\caption[...]{Contours of $\ln({\cal L})$ 
in steps of 1 relative to the
no-mixing case. All values are negative, i.e.\ the maximum is on the
line $\sin^22\Theta_0=0$. The spectral parameters were held fixed at
$E_{\rm b}=3\times10^{53}\,{\rm erg}$ and
$\langle E_{\overline{\nu}_e}\rangle=14\,{\rm MeV}$. The relative
$\overline\nu_\mu$ temperature $\tau$ was 2.0 (upper panel) and 1.4
(lower panel).
\label{Fig8}}
\end{figure}

Therefore, as in Sect.~IV.B we next take the opposite point of view
and assume that SN theory is roughly correct so that we should keep
$E_{\rm b}$ fixed at $3\times10^{53}\,\rm erg$. In the first analysis
we allow $T_{\overline\nu_e}$ to float for a fixed $\Delta m^2$ and
$\sin^22\Theta_0$. In Fig.~\ref{Fig7} we show the relevant contours of
the maximum likelihood relative to the no-mixing case. Again, shaded
areas correspond to a diminished maximum likelihood. As in
Fig.~\ref{Fig6} the maximum likelihood has an absolute maximum for
$\tau=2.0(1.4)$, $\sin^22\Theta_0=1$ and
$\log(\Delta m^2/{\rm eV^2})\approx -5.7(5.6)$  with
$\Delta\ln({\cal L}_{\rm max})\approx 1.4(1.1)$. A local maximum with
$\Delta\ln({\cal L}_{\rm max})\approx 0.8(0.3)$ is found for
$\sin^22\Theta_0\approx 0.8$ and
$\log(\Delta m^2/{\rm eV^2})\approx -5$. A similar effect occurred in
Fig.~\ref{Fig6} where the SN binding energy was also allowed to float.

Next, we hold both spectral characteristics fixed, to wit
$E_{\rm b}=3\times10^{53}\,\rm erg$ and $T_{\overline\nu_e}=4.7\,\rm
MeV$ which corresponds to the low end of the range of predicted
$\langle E_{\overline\nu_e}\rangle$ values given in Eq.~(\ref{mean}).
The contours of $\ln({\cal L})$ 
relative to the no-mixing case are shown
in Fig.~\ref{Fig8} in steps of 1, again with $\tau=2.0$ (upper panel)
and $\tau=1.4$ (lower panel). Note that all contours now represent
negative $\Delta\ln({\cal L})$, i.e.\ diminished likelihood values. If
we take the predicted SN parameters seriously we arrive at the same
conclusion as in Sect.~IV.B, namely that the no-mixing case is
favored.

\begin{figure}
\centering\leavevmode
\epsfxsize=2.8in
\epsfbox{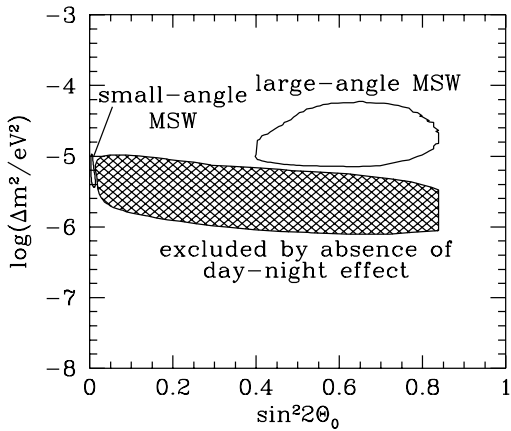}
\smallskip
\caption[...]{Mixing parameters favored by the MSW solutions of the
solar neutrino problem and those excluded by the absence of an
observed day-night effect at Kamiokande. (Contours according to Hata
and Haxton \cite{HataHaxton}.)
\label{Fig9}}
\end{figure}

Finally, we may suppose that future experiments will establish the
large-angle MSW solution of the solar neutrino problem, i.e.\ that the
mixing parameters lie within the indicated contour of Fig.~\ref{Fig9}.
Specifically, we choose the parameters $\sin^22\Theta_0=0.8$ and
$\Delta m^2=10^{-5}\,{\rm eV}^2$ with $\tau=1.0$, $1.4$, $1.7$, and
$2.0$ where $\tau=1.0$ corresponds to no mixing. As in Sect.~IV.B we
find that the 95.4\% confidence regions barely touch the lowest
predicted $\overline\nu_e$ energies only in the no-mixing case.
However, because of the Earth effect the other cases yield a serious
conflict only when the relative $\overline\nu_\mu$ temperature is
assumed to be large.

\begin{figure}
\centering\leavevmode
\epsfxsize=2.8in
\epsfbox{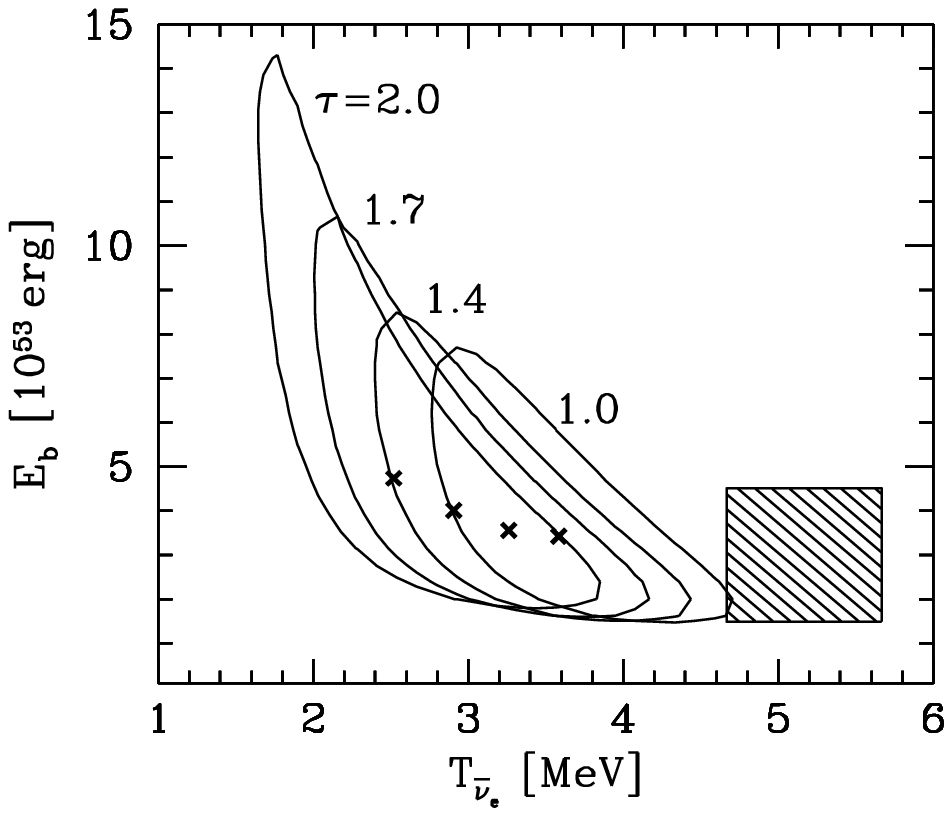}
\smallskip
\caption[...]{Best-fit values for $T_{\overline{\nu}_e}$ and
$E_{\rm b}$ and contours of constant likelihood which correspond to
95.4\% confidence regions. In each case a joint analysis between both
detectors was performed with $\sin^22\Theta_0=0.8$ and
$\Delta m^2=10^{-5}\,{\rm eV}^2$. The curves are marked with the
relative $\overline\nu_\mu$ temperature $\tau$. The hatched region
corresponds to the theoretical predictions of Eqs.~(\ref{E001}) and
(\ref{E001x}).
\label{Fig10}}
\end{figure}


\section{Discussion and Summary}

We have studied the impact of neutrino mixing on the interpretation of
the SN~1987A neutrino signal, focussing on those parameter regions
which are favored by the oscillation interpretation of the solar
neutrino problem. For these purposes the small-angle MSW solution is
equivalent to no mixing at all because only large vacuum mixing angles
lead to significant modifications of the antineutrino signal from a
SN. In agreement with previous authors we find that in the no-mixing
case the inferred neutron-star binding energy $E_{\rm b}$ and spectral
$\overline\nu_e$ temperature are consistent with theoretical
predictions, but only marginally so with regard to
$T_{\overline\nu_e}$; the 95.4\% confidence contour in the
$E_{\rm b}$-$T_{\overline\nu_e}$ plane just barely touches the
predicted range of average $\overline\nu_e$ energies given in
Eq.~(\ref{mean}).

\eject

Neutrino oscillation effects lead to a partial swap of the
$\overline\nu_e$ with the stiffer $\overline\nu_\mu$ spectrum. The
data already point to lowish neutrino energies, especially at the
Kamiokande detector, so that even a partial spectral swap aggravates
the disagreement between the predicted and experimentally inferred
neutrino energies.

For the large-angle MSW solution the regeneration effect in the Earth
always goes in the direction of partly undoing the swap caused by the
adiabatic oscillation in the SN envelope. Therefore, the 95.4\%
confidence contour in the $E_{\rm b}$-$T_{\overline\nu_e}$ plane may
be shifted only by a small amount, depending on the exact mixing
parameters, and depending on the relative $\overline\nu_\mu$
temperature (Fig.~\ref{Fig10}). Even for
$\tau=T_{\overline\nu_\mu}/T_{\overline\nu_e}=2.0$ it would be
difficult to claim a truly convincing conflict between observations
and SN theory. Of course, the true value of $\tau$ is not known. Put
another way, if the large-angle MSW solution would be borne out by
future solar neutrino experiments, the observed SN~1987A signal would
have to be taken as evidence for a soft $\overline\nu_\mu$ spectrum
relative to the $\overline\nu_e$ one.

The solar ``vacuum solution'' corresponds to a very small $\Delta m^2$
for which the SN oscillations are not adiabatic, i.e.\ we have vacuum
oscillations between the SN and here, and no regeneration effect in
the Earth. In this case the tension between the predicted and
observationally inferred SN neutrino spectra would be too significant
to ignore, i.e.\ one would be forced to take the possibility seriously
that the $\overline \nu_\mu$ spectra and/or $\overline\nu_e$ spectra
are softer than had been thought previously. Conversely, if one could
show that theoretical spectral predictions were accurate within the
claimed range of possibilities, then one would have to agree with the
findings of Smirnov, Spergel, and Bahcall \cite{SSB} that the solar
vacuum solution is incompatible with SN~1987A data. The conclusion of
Kernan and Krauss \cite{KK} that large mixing angles were actually
favored by the data can be upheld only if one ignores current 
theoretical
predictions of the SN spectra. In this case, indeed, the likelihood
function has a maximum for large mixing angles.

At the present time we would argue that the theoretical predictions of
SN neutrino spectra is not well enough established to achieve a
convincing selection between one of the three solutions of the solar
neutrino problem. We note, for example, that current numerical
calculations of the nonelectron-flavored neutrino spectra are based
on energy-conserving neutrino-nucleon scatterings between their energy
sphere and transport sphere in a SN core. However, nuclear recoils as
well as inelastic modes of energy transfer may soften these spectra in
a nonneglibile fashion \cite{JKRS}. There may be other novel effects
which modify these spectra.

Therefore, we believe that one should view the solar neutrino
experiments as one method for shedding new light on SN neutrino
spectra. Of course, the most interesting case would be if one of the
large-angle solutions would obtain as they would provide nontrivial
new information on the spectral characteristics of the SN~1987A
neutrinos.


\section*{Note Added}

After this paper had been submitted for publication, a new study has
appeared where the impact of gravitational fields on the phase
evolution of oscillating neutrinos is investigated \cite{gravitation}.
We believe that for the range of mixing parameters and oscillation
paths considered in our paper the gravitationally induced phases 
do not cause an observable effect.


\section*{Acknowledgments}

We thank H.-T.~Janka for numerous discussions concerning the predicted
SN neutrino spectra and for very helpful comments on the manuscript.
This research was supported, in part, by the European Union contract
CHRX-CT93-0120 and by the Deutsche Forschungsgemeinschaft grant
SFB~375.



\begin{references}

\bibitem{MS} S.~P.~Mikheev and A.~Yu.~Smirnov,
      Zh. Eksp. Teor. Fiz. {\bf 91}, 7 (1986)
      [Sov. Phys. JETP {\bf 64}, 4 (1986)].

\bibitem{Kam} K.~S.~Hirata, et al., Phys. Rev. Lett. {\bf 58}, 1490
      (1987) and Phys. Rev.~D {\bf 38}, 448 (1988).

\bibitem{IMB} R.~M.~Bionta et~al.,
      Phys. Rev. Lett. {\bf 58}, 1494 (1987).
      C.~B.~Bratton et~al., Phys. Rev.~D {\bf 37}, 3361 (1988).

\bibitem{Neutronization} D.~N\"otzold, Phys. Lett.~B {\bf 196}, 315
      (1987). J.~Arafune, M.~Fukugita, T.~Yanagida, and
      M.~Yoshimura, Phys. Lett.~B {\bf 194}, 477 (1987) and
      Phys. Rev. Lett. {\bf 59}, 1864 (1987). T.~P.~Walker and
      D.~N.~Schramm, Phys. Lett.~B {\bf 195}, 331 (1987).
      H.~Minakata, H.~Nunokawa, K.~Shiraishi, and H.~Suzuki, Mod.
      Phys. Lett.~A {\bf 2}, 827 (1987). H.~Minakata and H.~Nunokawa,
      Phys.  Rev.~D {\bf 38}, 3605 (1988). S.~P.~Rosen, Phys. Rev.~D
      {\bf 37}, 1682 (1988). T.~K.~Kuo and J.~Pantaleone, Phys. Rev.~D
      {\bf 37}, 298 (1988).

\bibitem{HataHaxton} N.~Hata and W.~Haxton,
       Phys. Lett. B {\bf 353}, 422 (1995).

\bibitem{Janka2} H.-T.~Janka, in: Proceedings Vulcano Workshop 1992,
       Frontier Objects in Astrophysics and Particle Physics, Conf.
       Proc. Vol.~40 (Soc. Ital. Fis.).

\bibitem{Janka1} H.-T.~Janka, in: Proceedings of the 5th Ringberg
       Workshop 1989 on Nuclear Astrophysics.
       K.~Sato and H.~Suzuki, Phys.~Lett.~B {\bf 196}, 267 (1987).

\bibitem{Wolfenstein} L.~Wolfenstein, Phys. Lett.~B {\bf 194}, 197
      (1987).
      P.~O.~Lagage, M.~Cribier, J.~Rich and D.~Vignaud,
      Phys. Lett.~B {\bf 193}, 127 (1987).

\bibitem{SSB} A.~Yu.~Smirnov, D.~N.~Spergel and J.~N.~Bahcall,
      Phys. Rev. D {\bf 49}, 1389 (1994).

\bibitem{Loredo89}  T.~J.~Loredo, D.~Q.~Lamb, Ann. N. Y. Acad. Sci.
      {\bf 571}, 601 (1989).

\eject

\bibitem{Loredo95} T.~J.~Loredo and D.~Q.~Lamb, Preprint,
      submitted to Physical Review D, 1995.

\bibitem{RS} G.~Raffelt and J.~Silk, Phys. Lett. B {\bf 366}, 429 
     (1996).

\bibitem{KP} V.~Barger, R.~J.~N.~Phillips, and K.~Whisnant,
      Phys. Rev.  Lett. {\bf 69}, 3135 (1992). P.~I.~Krastev and
      S.~T.~Petcov, Phys. Rev. Lett. {\bf 72}, 1960 (1994).

\bibitem{KK} P.~J.~Kernan and L.~M.~Krauss,
      Nucl. Phys.~B {\bf 437}, 243 (1995).

\bibitem{Smirnov} A.~Yu.~Smirnov, in: V.~A.~Kozyarivsky (ed.),
      Proceedings of the Twentieth International Cosmic Ray
      Conference, Moscow, 1987 (Nauka, Moscow, 1987).

\bibitem{Janka89}  H.-T.~Janka and W.~Hillebrandt,
      Astron. Astrophys. {\bf 224}, 49 (1989) and
      Astron. Astrophys. Suppl. {\bf 78}, 375 (1989).
      P.~M.~Giovanoni, P.~C.~Ellison, and S.~W.~Bruenn, Astrophys. J.
      {\bf 342}, 416 (1989).
      H.-T.~Janka, Neutrino Transport in Type~II Supernovae and
      Protoneutron Stars by Monte Carlo Methods, Ph.~D.~Thesis,
      Technische Universit\"at M\"unchen, MPA-Report No.~587 (1991).

\newpage

\bibitem{numericalmodels}
      S.~W.~Bruenn, Phys. Rev. Lett. {\bf 59}, 938 (1987).
      R.~Mayle, J.~R.~Wilson, and D.~N.~Schramm, Astrophys. J.
      {\bf 318}, 288 (1987).
      A.~Burrows, Astrophys. J. {\bf 334}, 891 (1988).
      E.~S.~Myra and A.~Burrows, Astrophys. J. {\bf 364}, 222 (1990).

\bibitem{Burrows88}  A.~Burrows, Astrophys. J. {\bf 334}, 891 (1988).

\bibitem{Janka95} H.-T.~Janka, Astropart. Phys. {\bf 3}, 377 (1995).

\bibitem{Eadie} W.~T.~Eadie, D.~Drijard, F.~E.~James, M.~Roos, and
      B.~Sadoulet, {\sl Statistical Methods in Experimental Physics\/}
      (North-Holland, Amsterdam, 1971).

\bibitem{Kendall} M.~G.~Kendall, A.~Stuart, {\sl The Advanced Theory
      of Statistics\/}, Vol.~2, 4th ed. (Griffin, London 1979).

\bibitem{JKRS} H.-T.~Janka, W.~Keil, G.~Raffelt, and D.~Seckel, Report
      ASTRO-PH/9507013, to be published in Physical Review Letters.

\bibitem{gravitation} D.~V.~Ahluwalia and C.~Burgard, E-print
      gr-qc/9603008 (1996), submitted to Physical Review Letters. 

\end{references}
\end{document}